\documentclass{appolbtr}

\usepackage{graphicx}
\usepackage{amsmath}
\usepackage{colortbl} 
\usepackage{color}
\usepackage{hyperref}
\hypersetup{colorlinks=false, linkcolor=black, filecolor=black, urlcolor=black}

\newcommand{\Fh}[2]{\,{}_#1F_#2}
\newcommand{\Fs}[3]{\!\!\left[\begin{array}{c}#1\,;\\#2\,;\end{array}#3\right]}
\newcommand{\Fz}[3]{\Fs{#1}{#2}{#3}}
\newcommand{\DD}{\frac{d}{2}}
\newcommand{\DDD}{\frac{d-1}{2}}
\newcommand{\DDDD}{\frac{d-2}{2}}

\allowdisplaybreaks
\newcommand{\bea}{\begin{eqnarray}} 
\newcommand{\eea}{ \end{eqnarray}} 
\newcommand{\ba}{\begin{eqnarray*}}
\newcommand{\ea}{  \end{eqnarray*}}

\newcommand{\beq}{\begin{equation*} }
\newcommand{\eeq}{\end{equation*}  }

\newcommand{\bqa}{\begin{eqnarray*} }
\newcommand{\eqa}{\end{eqnarray*}}

\begin{document}  \sloppy
\title{%
\vspace*{-20mm}
{\flushright{ 
\small \texttt{KW 17-002}
\\
\small \texttt{DESY 17-186}
\\
\small \texttt{DO-TH 17/29}
\\[5mm]
}}
Scalar one-loop vertex integrals as meromorphic functions of space-time 
dimension d
\thanks{Presented at workshop
 ``Matter To The Deepest'', XLI International Conference on
 Recent Developments in Physics of Fundamental Interactions
  (MTTD 2017), September 3-8, 2017, Podlesice, Poland, 
 \url{http://indico.if.us.edu.pl/event/4/overview}, to appear in the proceedings
}}
\author{%
\vspace{3mm}
Johannes Bl\"umlein%
\address{%
\vspace{-3mm}
Deutsches Elektronen-Synchrotron DESY, Platanenallee 6, 15738 Zeuthen, Germany}
\\[3mm] 
{Khiem Hong Phan%
\address{%
\vspace{-3mm}
Deutsches Elektronen-Synchrotron DESY, Platanenallee 6, 15738 Zeuthen, Germany
\\ 
and 
\\
University of Science, Vietnam National University, 
227 Nguyen Van Cu, Dist.5, Ho Chi Minh City,
Vietnam
}}
\\[3mm]
{Tord Riemann%
\thanks{Speaker}
\thanks{E-mail: tordriemann@gmail.com}}
\address{%
\vspace{-3mm}
Institute of Physics, University of Silesia, ul. 75 Pu\l{}ku Piechoty 1, 41-500 Chorz\'{o}w
\\ 
and 
\\
Deutsches Elektronen-Synchrotron, DESY, Platanenallee 6, 15738 Zeuthen, Germany}
}

\maketitle
\begin{abstract}
Representations are derived for the basic scalar one-loop vertex Feynman integrals as meromorphic functions of 
the  space-time dimension $d$
in terms of (generalized) hypergeometric functions $_2F_1$ and $F_1$. 
Values at asymptotic or exceptional kinematic points as well as  expansions around the singular points at 
$d=4+2n$, $n$ non-negative integers, may be derived from the representations easily. 
The Feynman integrals studied here may be used as building blocks for the calculation of one-loop and higher-loop scalar and 
tensor  amplitudes. 
From the recursion relation presented, higher $n$-point functions may be obtained in a straightforward manner.
\end{abstract}
  
\section{Introduction}
The systematic treatment of Feynman integrals is one of the basic ingredients of any perturbative calculation in quantum 
field theory.
In gauge field theories, the Feynman integrals may have both ultraviolet and infrared singularities, and the necessary regularizations 
are usually performed using a space-time dimension $d=4-2\epsilon$, where $\epsilon$ is the regulator.
At one loop, one has to treat two issues concerning dimensionally regularized Feynman integrals: 
(i) the calculation of $n$-point integrals;
(ii) the calculation of tensor integrals.
In a variety of publications it has been shown that, besides a direct calculation, a general $n$-point tensor Feynman integral may be 
algebraically reduced to a basis of scalar one- to four-point functions \cite{Davydychev:1991va}, with higher powers $\nu$ of 
propagators 
and in higher dimensions $d=4+2n-2\epsilon$.
Using recurrence relations \cite{Chetyrkin:1981qh,Tkachov:1981wb,Tarasov:1996br,Fleischer:1999hq}, one may get representations with all 
$\nu=1$, although yet at 
$d=4+2n-2\epsilon$ with non-negative integer $n$ . Such a representation in higher space-time dimensions may be organized such 
that it avoids the creation of inverse Gram determinants, which are known to destabilize realistic loop calculations 
\cite{Fleischer:2010sq,Riemann:2013acat}.

Having all this in mind it is evident that the seminal articles by 't Hooft and Veltman on scalar one-loop integrals 
\cite{'tHooft:1978xw} and  by Passarino and Veltman on tensor reduction \cite{Passarino:1978jh} for one- to four point functions in 
1978 set the 
stage for decades.
They solved the determination of the  Laurent-expansions in $\epsilon$ for these functions from the leading singular terms upto
including 
the constant terms, at $n=0$.
Later, the leading $\epsilon$-terms were determined in \cite{Nierste:1992wg}, and the general expansion in $\epsilon$ was studied 
in 
\cite{Passarino:2001wv}, again at $n=0$.

Although there are many attempts to determine the scalar one-loop integrals as meromorphic functions in the space-time dimension $d$, a 
complete solution in terms of special functions has not been given so far.
The most important article on the subject is \cite{Fleischer:2003rm}, where solutions have been found for scalar one- to 
four-point integrals in $d$ dimensions by solving iterative difference equations for them.
The solutions depend on Gauss' hypergeometric function for two-point functions, additionally on  the 
Appell function $F_1$ (a special case of the Kamp\'{e} de F\'{e}riet function and one of the set of Horn functions) for three-point 
functions,
and 
additionally on the Lauricella-Saran function $F_S$ for four-point functions.
In our understanding, the study \cite{Fleischer:2003rm} is not complete because the authors failed to determine sufficiently general 
expressions for certain boundary 
terms which they call $b_3$.

In this article, we close the  above-mentioned gap left in \cite{Fleischer:2003rm} by applying  another technique, starting from Feynman 
parameter representations for the 
Feynman integrals, deriving an iterative master integral.
For vertices, we solve here the iterative 
two-dimensional Mellin-Barnes representation.
Our version of the boundary term $b_3$ allows to cover the complete physical kinematics in the 
complex $d$-plane.  The most interesting case of four-point functions with a term $b_4$ has also been solved and will be published 
elsewhere.
  
\section{Definitions}
The scalar one-loop $N$-point Feynman integrals are defined as
\begin{eqnarray}\label{npoint}
J_n 
\equiv 
J_{n}(d; \{p_ip_j\}, \{m_i^2\})
 =
    \int \dfrac{d^d k}{i \pi^{d/2}} \dfrac{1}{D_1^{\nu_1} D_2^{\nu_2}\cdots D_n^{\nu_n}}
,
\end{eqnarray}
with inverse propagators $D_i= (k+q_i)^2-m_i^2+i\varepsilon$.
We assume $\nu_i=1$ as well as momentum conservation and all momenta to be incoming, 
$\sum_{i}^np_i=0.$ 
The $q_i$ are loop momenta shifts and will be expressed for applications by the external momenta $p_i$.
The $F$-function is independent of a shift of the integration variable $k$ due to the dependence on the differences $q_i-q_j$.
Further, the difference of two neighboring momentum shifts $q_i$ equals to an external momentum.
We use the Feynman parameter representation for the evaluation of the Feynman integrals \eqref{npoint}:
\begin{eqnarray}\label{bh-1}
J_n
&=& 
(-1)^{n}
{\Gamma\left(n -d/2\right)}
   \int_0^1 \prod_{j=1}^n dx_j
\delta \left( 1-\sum_{i=1}^n x_i \right)
\frac { 1 } {F_n(x)^{n-d/2}}
.
\end{eqnarray}
Here, the $F$-function is the second Symanzik polynomial. It is  derived from the propagators, 
$M^2 \equiv {x_1D_1+   \cdots + x_n D_n} ~=~ k^2 - 2Qk + J.$
Using  $\delta (1-\sum x_i)$ under the integral in order to transform linear terms in $x$ into bilinear ones, one obtains
\begin{eqnarray}
F_n(x) &=&  -\left(\sum_{i=1}^n x_i\right)\times J +Q^2 ~=~\frac{1}{2} \sum_{i,j} x_i Y_{ij} x_j-i\varepsilon
,
\end{eqnarray}
where the $Y_{ij}$ are elements of the Cayley matrix, introduced for a systematic study of one-loop $n$-point Feynman integrals e.g. in 
\cite{Melrose:1965kb}:
\begin{eqnarray}\label{eq-yij}
Y_{ij} = Y_{ji} &=& m_i^2+m_j^2-(q_i-q_j)^2 . 
\end{eqnarray}
We will discuss the one-loop integrals as functions of two kinematic matrices and determinants, which were introduced by Melrose
\cite{Melrose:1965kb}. 
The Cayley determinant $\lambda_{12\dots n}$ is composed of the $Y_{ij}$ introduced in \eqref{eq-yij}, 
and its determinant is:
\begin{eqnarray} \label{npoint-cayley}
\lambda_n \equiv \lambda_{12 \dots n}
&=&  
\left|
\begin{array}{cccc}
Y_{11}  & Y_{12}  &\ldots & Y_{1n} \\
Y_{12}  & Y_{22}  &\ldots & Y_{2n} \\
\vdots  & \vdots  &\ddots & \vdots \\
Y_{1n}  & Y_{2n}  &\ldots & Y_{nn}
\end{array}
         \right|
.
\end{eqnarray}
We also define the $(n-1)\times(n-1)$ dimensional Gram determinant,
\begin{eqnarray}
 \label{Gram}
G_n \equiv G_{12 \cdots n}
= 
- 
\left|
\begin{array}{cccc}
  \! (q_1-q_n)^2 
&\ldots & (q_1-q_n)(q_{n-1}-q_n) 
\\
  \! (q_1-q_n)(q_2-q_n) 
&\ldots 
& (q_2-q_n)(q_{n-1}-q_n) 
\\
  \vdots    
&\ddots   & \vdots 
\\
  \! (q_1-q_n)(q_{n-1}-q_n)    
&\ldots & (q_{n-1}-q_n)^2
\end{array}
\right|.
\end{eqnarray}
Both determinants are independent of a common shifting of the momenta $q_i$.
After elimination of one $x$-variable from the $n$-dimensional integral \eqref{npoint}, e.g. $x_n$, by use of the $\delta$-function in 
\eqref{bh-1},
the $F$-function becomes a quadratic form in $x=(x_i)$ with linear terms in $x$ and with an inhomogeneity $R_n$:
\begin{eqnarray} 
F_n(x)=
 (x - y)^T G_n (x-y)
 + r_{n} { ~ - i \varepsilon} =  \Lambda_{n}(x) +R_{n}
.
\end{eqnarray}
The following relations are also valid:
\begin{eqnarray}
R_n
\equiv r_n  - i \varepsilon
= -~\frac{\lambda_n}{G_n} - i \varepsilon
\end{eqnarray}
and
\begin{eqnarray} \label{eq-def-yi}
y_i &=& \dfrac{\partial r_{n}}{\partial m_i^2} 
~=~   - \frac{1}{G_{n}}~\dfrac{\partial \lambda_n}{\partial m_i^2} 
~\equiv~ - \dfrac{\partial_i \lambda_n}{g_{n}}
 , 
~~~i=1 \cdots n.
\end{eqnarray} 
The auxiliary condition $\sum_i^n y_i =1$ is fulfilled.
The notations for the $F$-function are finally independent of the choice of the variable which was eliminated by use of the 
$\delta$-function in the integrand of \eqref{bh-1}.  
The inhomogeneity $R_{n}$ 
is the only variable carrying the causal $i\varepsilon$-prescription, while e.g. $\Lambda(x)$ and the $y_i$ are by definition real.

The simplest case of a one-loop scalar Feynman integral is the one-point function or tadpole,
\begin{eqnarray} \label{eq-tadpole}
J_{1}(d;m^2) 
&=& 
\int \dfrac{d^d k}{i \pi^{d/2}} 
\dfrac{1}{k^2-m^2+i \varepsilon}
~~=~~
- \frac{\Gamma( 1 -d/2)}
 {(m^2-i \varepsilon)^{1-d/2}}
.
\end{eqnarray}

Finally, we introduce the operator ${\bf k^{-}}$, which 
will reduce an $n$-point Feynman integral $J_n$ to an $(n-1)$-point integral $J_{n-1}$ 
by shrinking the $k^{th}$ propagator, $1/D_k$:
\begin{eqnarray}\label{eq-def-operator-k-}
{\bf k^{-}}~ J_n 
&=& 
{\bf k^{-}} \int \dfrac{d^d k}{i \pi^{d/2}} \dfrac{1}{\prod_{j=1}^n D_j}
~~=~~
\int \dfrac{d^d k}{i \pi^{d/2}} \dfrac{1}{\prod_{j\neq k,j=1}^n D_j}
.
\end{eqnarray}

\section{The master formula for the Feynman integrals $J_n$ 
\label{sec-one-one}}
We study the general case with $G_{n} \neq 0$ 
and $R_{n} \neq 0$. Other cases are simply derived from the formulae given here.
One may use the well-known Mellin-Barnes representation in order to decompose the 
integrand of $J_n$ given in \eqref{bh-1} as follows:
\begin{eqnarray}
\label{c5}
\dfrac{1}{[\Lambda_n(x) + R_n]^{ n-\frac{d}{2} }}
= \dfrac{R_n^{-(n- \frac{d}{2})}}{2\pi i} 
\int\limits_{-i\infty}^{+i\infty}ds \; 
  \dfrac{\Gamma(-s)\;\Gamma( n-\frac{d}{2} +s) } { \Gamma(n-\frac{d}{2}) }    
  \left[\dfrac{\Lambda_n(x)}{ R_n} \right]^s,
\end{eqnarray}
for 
$|\mathrm{Arg}(\Lambda_n/R_n)|<\pi$. 
The condition always applies.
As a result of \eqref{c5}, the Feynman parameter integral of $J_n$ becomes homogeneous:
\begin{eqnarray}
\label{c6}
K_n = \prod_{j=1}^{n-1} \int_0^{1-\sum_{i=j+1}^{n-1} x_i}  dx_j
\; 
\left[ 
      \dfrac{\Lambda_{n}(x)}{ R_n} \right]^s  ~\equiv~
\int dS_{n-1}
\; 
\left[ 
      \dfrac{\Lambda_{n}(x)}{ R_n} \right]^s 
.
\end{eqnarray}
 In order to solve this integral, we introduce the differential 
operator $\hat{P}_n$ \cite{Bernshtein1971-from-springer,Golubeva:1978},
\begin{eqnarray}
\label{c7}
\frac{\hat{P}_n}{s} \left[\dfrac{\Lambda_n(x)}{R_n} \right]^s
~\equiv~ 
\sum_{i=1}^{n-1} 
   \frac{1}{2s} (x_i-y_i)\frac{\partial}{\partial x_i}
\left[\dfrac{\Lambda_n(x)}{R_n} \right]^s
   =    \;\left[\dfrac{\Lambda_n(x)}{R_n } \right]^s
,
\end{eqnarray}
into the integrand of \eqref{c6}:
\small 
\begin{eqnarray}
\label{c8}
\nonumber
K_n
&=& \frac{1}{s}
\int dS_{n-1}\; \hat{P}_n 
    \left[ \dfrac{\Lambda_n(x)}{R_n} \right]^s
~=~ \frac{1}{2s} \sum_{i=1}^{n-1} 
\prod_{k=1}^{n-1}
\int\limits_0^{u_k}  dx'_k ~ 
   (x_i-y_i)\frac{\partial}{\partial x_i} 
   \left[\dfrac{\Lambda_n(x)}{R_n} \right]^s
.
\end{eqnarray}
\normalsize 
\noindent 
After a series of manipulations in order to perform one of the $x$-integrations -- by partial integration, eating the 
corresponding differential --, and applying a Barnes relation \cite{Barnes:zbMATH02640947} 
(item 14.53 at page 290 of~\cite{watson}), 
one  arrives at the following recursion relation:
\begin{eqnarray} \label{JNJN1}
J_n(d,\{q_i,m_i^2\})
&=& \dfrac{{-1}}{2\pi i} \int\limits_{-i\infty}^{+i\infty}ds  
     \dfrac{\Gamma(-s) \Gamma(\frac{d-n+1}{2}+s) \Gamma(s+1) }
           { 2\Gamma(\frac{d-n+1}{2}) } 
     \left(\frac{1}{{R_{n}}}\right)^s
\nonumber\\
&&\hspace{0.4cm}\times     \sum\limits_{k=1}^n 
     \left( \frac{1}{{R_{n}}} 
            \frac{\partial {r_n}}{\partial m_k^2} 
      \right) \;
     {\bf k}^- J_n (d+2s;\{q_i,m_i^2\}).
\end{eqnarray}
Eq. \eqref{JNJN1}  is the master integral for one-loop $n$-point functions in space-time dimension $d$, representing 
them by $n$ 
integrals over $(n-1)$-point functions with a shifted dimension $d+2s$.
This Mellin-Barnes integral representation is the equivalent  to Eq.~(19) of ~\cite{Fleischer:2003rm}. 
There, an {\it infinite sum} over a discrete parameter $s$ was derived in order to represent an $n$-point function in space-time dimension 
$d$ by simpler functions $J_{n-1}$ at dimensions $d+2s$.

\section{The three-point function} 
According to the master formula \eqref{JNJN1}, we can write the massive 3-point function as a sum of three terms, each of them relying on a 
two-point function, relying on one-point functions.
After analytically performing the two-fold Mellin-Barnes integrals, we arrive at: 
\small 
\begin{eqnarray}
\label{J3normal}
J_3(d; \{p_i^2\}, \{m_i^2\})   &=& J_{123} + J_{231} + J_{312}, 
\end{eqnarray}
with 
\begin{eqnarray}
\label{J123normal}
J_{123} &=&  
   \Gamma\left( 2-\DD\right)
~ R_{3}^{\DD-2}  
~ b_{123}
\nonumber\\[3mm]
&& - ~~ \dfrac{\sqrt{\pi}~\Gamma\left( 2-\DD\right) 
   \Gamma\left(\DD-1\right)  }{ \Gamma\left(\frac{d-1}{2}\right) }
   ~~ \dfrac{\partial_3 \lambda_{3}}{\lambda_{3}} 
~~
 \dfrac{ R_{12}^{\DD-1} }{4\lambda_{12}} 
   \left[  \dfrac{\partial_2 \lambda_{12}}{\sqrt{1-\frac{m_1^2}{R_{12}}}}
  + \dfrac{\partial_1 \lambda_{12} }{\sqrt{1-\frac{m_2^2}{R_{12}}}} \right]
\nonumber\\
&& \times ~  \Fh21\Fz{ \DDDD, 1}{\DDD}{ \dfrac{R_{12}}{R_{3} } }                     
~
{+} 
~\frac{2}{d-2}\Gamma\left( 2-\DD\right)
~ \dfrac{\partial_3 \lambda_{3}}{\lambda_{3}}
\\[3mm]
&& \hspace{1cm} 
\times 
   \left[
\dfrac{\partial_2 \lambda_{12} }{\sqrt{1-\frac{m_1^2}{R_{12}}} }
   \dfrac{(m_1^2)^{\frac{d}{2} -1 } }{4\lambda_{12}} 
   F_1 \left(\dfrac{d-2}{2}; 1, \frac{1}{2}; \DD; 
             \frac{m_1^2}{R_{3}}, \dfrac{m_1^2}{R_{12}} \right) 
+  (1\leftrightarrow 2)      
\right],  \nonumber
\end{eqnarray}
\normalsize\noindent
and 
\small 
\begin{eqnarray}
\label{b123normal} 
b_{123} &=&  -\frac{1}{2G_{12}}
    ~
 \dfrac{\partial_3 \lambda_{3}}{\lambda_{3}} 
~
    \left(  \dfrac{\partial_2 \lambda_{12}}{\sqrt{1-\frac{m_1^2}{R_{12}}}}
  + \dfrac{\partial_1 \lambda_{12} }{\sqrt{1-\frac{m_2^2}{R_{12}}}} \right)
    \Fh21\Fz{1, 1}{\frac{3}{2}}{ \dfrac{R_{12}}{R_{3} } }                     \\
&&- \dfrac{\partial_3 \lambda_{3}}{\lambda_{3}}
    \left \{ \dfrac{\partial_2 \lambda_{12}}{\sqrt{1-\frac{m_1^2}{R_{12}}}} 
    \; \dfrac{ m_1^2 }{4\lambda_{12}} \;
    F_1 \left(1; 1, \frac{1}{2}; 2; 
              \frac{m_1^2}{R_{3}}, \dfrac{m_1^2}{R_{12}} \right)
    +  (1\leftrightarrow 2)      \right \},
\nonumber
\end{eqnarray}
\normalsize
\noindent
where $\partial_i \lambda_{j\cdots}$ is defined in \eqref{eq-def-yi}.
The representation \eqref{J3normal} is valid for
 Re$\Big(\frac{d-2}{2} \Big)>0$.
The conditions $\Big|\frac{m_{i}^2}{R_{ij}} \Big|<1$, $\Big|\frac{R_{ij}}{R_{3}} \Big|<1$ had to be met during the derivation. The 
result 
may be analytically continued in a straightforward way, however, in the complete complex domain.
\section{Vertex numerics}
In Table \ref{table-j3table7line1} we show one numerical case, further examples are given in the slides of the presentation at MTTD 
2018, see  \url{http://indico.if.us.edu.pl/event/4/}.
While we agree completely with the ``main'' parts of the solutions for the Feynman integrals given in \cite{Fleischer:2003rm}, our 
boundary term has a richer structure and is, contrary to $b_3$ \cite{Fleischer:2003rm},  valid for arbitrary kinematics without 
additional specific considerations.

\begin{table}[t]
\begin{center}
\begin{tabular}{|l|l|l|}\hline \hline
      [$p_i^2$],~~[$m_i^2$]    &   [-100, {\bf {+200}}, -300],~~~[10, 20, 30]  
\\ \hline
$G_{3}$,  $\lambda_{3}$ &{\bf {+480000}}, --19300000
\\
$m_i^2/r_{3}$  & 0.248705, 0.497409, 0.746114
\\
$\sum J$, eq. \eqref{J123normal} &
--0.012307377 -- 0.056679689 I
\\ \hline \hline
$\sum b$, eq. \eqref{b123normal} &
 + 0.047378343 I 
\\
$J_3^{(TR)}$ $= \sum J+\sum b$ &
--0.012307377 -- 0.009301346 I
\\ \hline  \hline 
$b_3$ \cite{Fleischer:2003rm} &
 + 0.047378343 I 
\\
$b_3$+$\sum J~ \cite{Fleischer:2003rm}$ &
--0.012307377 -- 0.009301346 I
\\ \hline
$J_3^{(OT)}=\sum J$ &
 {$b_3\to 0$, {\bf gets wrong}} 
\\ \hline \hline
(-1)$\times$FIESTA 3 \cite{Smirnov:2013eza}&
--(0.012307 + 0.009301 I)
\\ \hline \hline
LoopTools/FF \cite{Hahn:1998yk} &--0.012307377 -- 0.009301346 I 
\\ \hline\hline
\end{tabular}
\end{center}
\caption[]{
\label{table-j3table7line1}
\footnotesize
Numerics for the constant term of a vertex in space-time dimension  $d=4-2\epsilon$. Causal $\varepsilon =   10^{-20}$. This work, 
\eqref{J3normal} to \eqref{b123normal} is 
labelled (TR).
Bold input quantities suggest that, according to eq. 
(73) in \cite{Fleischer:2003rm} (labelled (OT)), one has to set there $b_3=0$.
This choice gives a wrong result for $J_3$. If instead we choose in the numerics for eq. (75) of \cite{Fleischer:2003rm} that 
$\sqrt{-G_{3}}\to \sqrt{-G_{3}+ \varepsilon I} = +I\sqrt{|G_3|}$, and include the non-vanishing value for $b_3$, the $J_3^{(OT)}$ gets 
correct. The 
setting $G_3-\varepsilon I$
 looks counter-intuitive for a ``momentum''-like function like $G_3$. 
}
\end{table}

 \section*{Acknowledgements}
This work was supported in part by the European Commission through contract 
PITN-GA-2012-316704 ({HIGGSTOOLS}). 
K.H.P. would like to thank DESY for hospitality and support during the present project.
His work is also supported by the Vietnam National Foundation for Science
and Technology Development (NAFOSTED) under the grant No 103.01-2016.33. 
The work of T.R. is supported in part by the Polish Alexander von Humboldt Honorary Research Fellowship 2015.
We thank Ievgen Dubovyk and Johann Usovitsch for assistance in several numerical comparisons with the MB-suite 
AMBRE/MB/mbtools/MBnumerics/CUBA \cite{Gluza:2007rt,Gluza:2010rn,Dubovyk:2016ocz,Czakon:2005rk,mbtools,Usovitsch:201511,Hahn:2004fe}.

\end{document}